\documentclass[aps,pre,onecolumn,superscriptaddress,notitlepage]{revtex4-1}
\usepackage{graphicx}% Include figure files
\usepackage{dcolumn}% Align table columns on decimal point
\usepackage{bm}% bold math
\usepackage{accents}
\usepackage{amssymb}
\usepackage{feynmp}
\usepackage{url}
\usepackage{color}
\usepackage{setspace}

\newcommand{\beq}{\begin{equation}}
\newcommand{\eeq}{\end{equation}}
\newcommand{\beqa}{\begin{eqnarray}}
\newcommand{\eeqa}{\end{eqnarray}}
\newcommand{\Tr}{\text{Tr}}

\newcommand{\av}[1]{\left\langle #1 \right\rangle}

\begin{document}

%\preprint{}

\title{Speed limit for open quantum systems}


\author{Ken Funo}
\email[]{ken.funo@riken.jp}
%\thanks{}
\affiliation{Theoretical Physics Laboratory, RIKEN Cluster for Pioneering Reserach, Wako-shi, Saitama 351-0198, Japan}
\author{Naoto Shiraishi}
\affiliation{Department of Physics, Keio university, Hiyoshi 3-14-1, Kohoku-ku, Yokohama, Japan}
\author{Keiji Saito}
\affiliation{Department of Physics, Keio university, Hiyoshi 3-14-1, Kohoku-ku, Yokohama, Japan}



\date{\today}

\begin{abstract}
We study the quantum speed limit for open quantum systems described by the Lindblad master equation. The obtained inequality shows a trade-off relation between the operation time and the physical quantities such as the energy fluctuation and the entropy production. We further identify a quantity characterizing the speed of the state transformation, which appears only when we consider the open system dynamics in the quantum regime. When the thermal relaxation is dominant compared to the unitary dynamics of the system, we show that this quantity is approximated by the energy fluctuation of the counter-diabatic Hamiltonian which is used as a control field in the shortcuts to adiabaticity protocol. We discuss the physical meaning of the obtained quantum speed limit and try to give better intuition about the speed in open quantum systems.  
\end{abstract}

\maketitle

\section{Introduction}
Quantum speed limit (QSL) is an inequality relation which sets a lower bound on the operation time required to transform a given initial state to a given final state~\cite{DC}. For an isolated quantum system, the Mandelstam-Tamm type QSL is given by $\tau\geq \hbar\mathcal{L}(\rho(0),\rho(\tau))/\langle\Delta E\rangle_{\tau}$, where $\mathcal{L}$ is the Bures distance between the initial state $\rho(0)$ and the final state $\rho(\tau)$ and $\langle\Delta E\rangle_{\tau}$ is the time-averaged value of the uncertainty in energy~\cite{MT,Fle,AA,Braunstein,Uhlmann,ML}. This inequality shows a trade-off between the operation time $\tau$ and the energy fluctuation, since the QSL has its origin in formulating the Heisenberg time-energy uncertainty relation. The QSL gives us important insight about the speed of the state transformation. In this case, large energy fluctuation is required to speedup the operation. Later studies have revealed the connection between the QSL and the geometry of a quantum state~\cite{AA,Braunstein,Uhlmann}. The QSL is independent of the details of the dynamics and can be applied to a wide range of dynamics, i.e., isolated and open quantum systems~\cite{Taddei13,Campo13,Deffner13,Zhang14,Pires16,Campaioli18a,Campaioli18b,Deffner17} and even for classical systems~\cite{Okuyama18,Shanahan18,Ito18,Shiraishi18}. Since the QSL is quite universal, it has many applications in the studies of quantum computation~\cite{Lloyd}, quantum metrology~\cite{Alipour14}, quantum optimal control~\cite{Santos15,Campbell17,Mukherjee13} and quantum thermodynamics~\cite{Funo}.

The most general form of the quantum speed limit can be obtained solely from the property of the geometry of a quantum state as follows. By introducing any kind of distance $D(\rho(0),\rho(\tau))$ between density matrices, the triangle inequality gives $ \int^{\tau}_{0}dt \sqrt{g_{tt}} \geq D(\rho(0),\rho(\tau))$, where $g_{tt}$ is a metric on the space of density matrices induced by $D$~\cite{Provost80}. 
By introducing the time-average as $\av{f}_{\tau}=\tau^{-1}\int^{\tau}_{0}dt f(t)$, we obtain the following formal inequality relation 
\beq
\tau\geq \frac{D(\rho(0),\rho(\tau))}{\av{\sqrt{g_{tt}}}_{\tau}}, \label{GOMSL}
\eeq
which is a geometric formulation of the QSL, and similar arguments apply to classical systems as well~\cite{Okuyama18,Shanahan18}. Note that Eq.~(\ref{GOMSL}) has been discussed for contractive Reimanninan metrics in Ref.~\cite{Pires16} and for Shatten-$p$ distances including the trace distance in Ref.~\cite{Deffner17}. Here, the metric $g_{tt}$ measures the quadratic sensitivity of the density matrix when varying $t$, and includes information about the generator $\mathcal{G}_{t}$ of the time-evolution. Therefore, we interpret the denominator of~(\ref{GOMSL}) as the average velocity. In particular, when the distance is given by $D(\rho,\sigma)=||\rho-\sigma||_{*}$, where $||X||_{*}$ is some norm of an operator $X$, the velocity term is given by the norm of the generator, i.e., $\sqrt{g_{tt}}=||\dot{\rho}||_{*}=||\mathcal{G}_{t}[\rho]||_{*}$.

We find from Eq.~(\ref{GOMSL}) that the formulation of the QSL depends on the choice of the distance between quantum states~\cite{Deffner17}. From a practical point of view, this arbitrariness can be utilized to find the QSL~(\ref{GOMSL}) which gives the tightest bound on the operation time $\tau$ for some specific settings~\cite{Pires16,Campaioli18a,Campaioli18b}. 
%From a practical point of view, one can consider different choices of the distance and try to find the QSL~(\ref{GOMSL}) which gives the tightest bound on the operation time $\tau$~\cite{Pires16,Campaioli18a,Campaioli18b}. It is also important to find the norm such that the velocity term $\sqrt{g_{tt}}$ can be measured experimentally~\cite{Campaioli18b}.
It is also important to find the distance such that the velocity term $\sqrt{g_{tt}}$ can be measured experimentally~\cite{Campaioli18b}.

From a fundamental point of view, the QSL gives us physical intuition about the speed of the state transformation. Therefore, the task is to relate the velocity term $\sqrt{g_{tt}}$ with a quantity which is physically meaningful. For an isolated quantum system, the velocity term can be related to the energy fluctuation as is the case for the Mandelstam-Tamm type QSL~\cite{MT}. In Ref.~\cite{Shiraishi18}, we have related the velocity term to the quantities that appear in stochastic thermodynamics and derived the speed limit for classical stochastic processes. However, for open quantum systems, most of the QSLs derived in the literatures remain as a formal mathematical expression, and physical intuition is hard to extract from the obtained results.

In this paper, we explore the physical meaning of the velocity term in open quantum systems described by the Lindblad master equation~\cite{Breuer,Albash12,Yuge17} and formulate the QSL. The obtained inequality consists of three velocity terms, where two of them coincide with the previously obtained QSLs~\cite{MT,Shiraishi18}. The first term is the energy fluctuation which characterizes the velocity for a unitary time-evolution generated by the system Hamiltonian. The second term is a combination of the entropy production and the dynamical activity~\cite{LAW,Gar,Maes1,Maes2,SST}, which characterize the thermodynamic irreversibility and the frequency of the jump processes in stochastic thermodynamics, respectively. The third term is relevant only when we consider the open system dynamics in the quantum regime, and has not been reported in previous studies. In the parameter regime where the thermal relaxation becomes dominant, we further relate this term with the energy fluctuation of the counter-diabatic Hamiltonian which is used as a control field to achieve the quantum adiabatic dynamics in a finite protocol time~\cite{STAR,DR03,DR05,Berry09}. The obtained QSL therefore provides us better intuition about the speed in open quantum systems.

This paper is organized as follows. In Sec.~\ref{sec:setup}, we explain the setup of our paper by introducing the Lindblad quantum master equation. We also introduce the quantum entropy production which is used in the main result. In Sec.~\ref{sec:derivation}, we derive our main result, the QSL for open quantum systems, by decomposing the generator of the time-evolution into three parts and bounding from above the norm of each part. In Sec.~\ref{sec:limit}, we consider three limiting cases. In the weak dissipation limit and the classical limit, we reproduce the previously obtained results. When the thermal relaxation is dominant, we give further physical explanation of the velocity term in the QSL by relating it to a quantity used in the control technique known as the shortcuts to adiabaticity. We summarize our result in Sec.~\ref{sec:summary}.

\section{\label{sec:setup}Setup}
\subsection{Lindblad master equation}
We consider an externally driven system interacting with the heat bath, described by the following Lindblad type master equation~\cite{Breuer,Albash12,Yuge17}
\beq
\partial_{t}\rho(t)=\mathcal{L}_{t}[\rho(t)]=-\frac{i}{\hbar}[H(t)+H_{\rm{LS}}(t),\rho(t)]+\mathcal{D}[\rho(t)], \label{LE}
\eeq
where $H(t)$ is the Hamiltonian of the system. In this paper, we consider a Hamiltonian which does not contain time-reversal symmetry breaking terms, such as magnetic fields and Coriolis force, and so on. We note that the Lamb shift term $H_{\rm{LS}}(t)$ induced by the bath has the effect of shifting the energy level of the system Hamiltonian $H(t)$, i.e., $[H(t),H_{\rm{LS}}(t)]=0$~\cite{Breuer}. For simplicity, we neglect the effect of the Lamb shift, but the main result is still true if we use the renormalized system Hamiltonian $H_{\rm{r}}(t)=H(t)+H_{\rm{LS}}(t)$. The term $\mathcal{D}[\rho(t)]$ is a quantum dissipater described as
\beq
\mathcal{D}[\rho(t)]=\sum_{\omega_{t},\alpha}\gamma_{\alpha}(\omega_{t})\Bigl[L_{\omega_{t},\alpha}(t) \rho(t)L^{\dagger}_{\omega_{t},\alpha}(t)-\frac{1}{2}\{ L^{\dagger}_{\omega_{t},\alpha}(t)L_{\omega_{t},\alpha}(t), \rho(t)\}\Bigr] .
\eeq
Here, $\{A,B\}=AB+BA$ is the anti-commutator. The Lindblad operator $L_{\omega_{t},\alpha}$ is given by
\beq
L_{\omega_{t},\alpha}(t)=\sum_{\omega_{t}=\epsilon_{m}(t)-\epsilon_{n}(t)}|\epsilon_{n}(t)\rangle\langle\epsilon_{n}(t)|L_{\alpha}|\epsilon_{m}(t)\rangle\langle \epsilon_{m}(t)| .
\eeq
This operator describes a quantum jump from one energy eigenstate $|\epsilon_{m}(t)\rangle$ to another $|\epsilon_{n}(t)\rangle$ with their energy difference equal to $\omega_{t}$ and thus satisfies 
\beq
[L_{\omega_{t},\alpha}(t),H(t)]=\omega_{t} L_{\omega_{t},\alpha}(t) , \label{Lcommute}
\eeq
and $L_{-\omega_{t},\alpha}(t)=L^{\dagger}_{\omega_{t},\alpha}(t)$. We further assume the detailed balance condition 
\beq
\gamma_{\alpha}(-\omega_{t})=\gamma_{\alpha}(\omega_{t})e^{-\beta\omega_{t}}, \label{detailed}
\eeq
which is a sufficient condition to make the Gibbs state $\rho_{\rm{eq}}(t)=\exp(-\beta H(t))/Z(t)$ to become the instantaneous stationary solution of Eq.~(\ref{LE}), 
i.e., $\mathcal{L}_{t}[\rho_{\rm{eq}}(t)]=0$. Here, $\beta$ is the inverse temperature of the heat bath.

\subsection{Quantum entropy production}
We utilize the notion of quantum stochastic thermodynamics and introduce the quantum entropy production which plays a central role in the main result. The quantum entropy production rate is defined as
\beq
\dot{\sigma}:=\dot{S}-\beta\dot{Q}, \label{EPrate}
\eeq
where
\beq
\dot{S}:=-\Tr[(\partial_{t}\rho(t))\ln\rho(t)]
\eeq
is the von Neuman entropy flux of the system %, $\beta$ is the inverse temperature of the bath, 
and 
\beq
\dot{Q}:=\Tr[(\partial_{t}\rho(t))H(t)]
\eeq
is the heat flux that is transfered from the bath to the system. Since $-\beta\dot{Q}$ can be interpreted as the entropy flux produced by the bath, $\dot{\sigma}$ quantifies the rate of the entropy variation of the entire system. The entropy production is one of the most fundamental quantity in stochastic thermodynamics since it quantifies the irreversibility of the thermodynamic process and satisfies the exact nonequilibrium relation called the fluctuation theorem~\cite{Seifert12,Esposito09,Campisi11,Horowitz,Hekking,Funoreview}. In what follows, we use the Lindblad master equation~(\ref{LE}) and rewrite Eq.~(\ref{EPrate}) to show the nonnegativity of $\dot{\sigma}$, i.e., the second law of thermodynamics. 

Let us first introduce the eigenbasis set $\{|n(t)\rangle\}$ which diagonalizes the density matrix of the system at time $t$, i.e., $\rho(t)=\sum_{n}p_{n}(t)|n(t)\rangle\langle n(t)|$. Readers should note that $|n(t)\rangle$ is different from the eigenstate of the Hamiltonian $|\epsilon_{n}(t)\rangle$. In the following, we omit the time dependence to simplify the notation. By introducing the quantity  
\beq
W^{\omega,\alpha}_{mn}:=\gamma_{\alpha}(\omega)| \langle m|L_{\omega,\alpha}|n\rangle|^{2}, \label{QMtrans}
\eeq
%to simplify the expression of $\dot{\sigma}$. 
%The amount of heat flux that is transfered from the bath to the system is given by
%\beq
%\dot{Q}:=\Tr[(\partial_{t}\rho(t))H(t)]=-\sum_{\omega_{t},\alpha}\gamma_{\alpha}(\omega_{t})\Tr[L_{\omega_{t},\alpha}(t)\rho(t)L^{\dagger}_{\omega_{t},\alpha}(t)]\omega_{t}. \label{OpenQ}
%\eeq
%Now the entropy flux produced by the bath can be expressed as
the entropy flux produced by the bath can be expressed as
\beqa
-\beta\dot{Q}&=&\sum_{\omega,\alpha}\gamma_{\alpha}(\omega)\Tr[L_{\omega,\alpha}\rho L^{\dagger}_{\omega,\alpha}]\beta\omega \nonumber \\
&=&\sum_{\omega,\alpha,n,m}\hspace{-3mm}{}^{'} W^{\omega,\alpha}_{mn}p_{n} \beta\omega \nonumber \\
&=&\sum_{\omega,\alpha,n,m}\hspace{-3mm}{}^{'} W^{\omega,\alpha}_{mn}p_{n} \ln\frac{W^{\omega,\alpha}_{mn}}{W^{-\omega,\alpha}_{nm}} , \label{dotQdet}
\eeqa
where $\sum \hspace{-0.5mm}{}^{'}$ denotes the summation by excluding indicies satisfying $(m=n)\wedge (\omega=0)$. %, or equivalently, $W^{\omega,\alpha}_{mn}p_{n}=W^{-\omega,\alpha}_{nm}p_{m}$. 
In Eq.~(\ref{dotQdet}), we use Eq.~(\ref{Lcommute}) and Eq.~(\ref{detailed}) and obtain the first and the third line, respectively. 
Similarly, we have
\beqa
\dot{S}&=&-\sum_{\omega,\alpha}\gamma_{\alpha}(\omega)\Tr[L_{\omega,\alpha}\rho L_{\omega,\alpha}^{\dagger}\ln\rho]+\sum_{\omega,\alpha}\gamma_{\alpha}(\omega)\Tr[L_{\omega,\alpha}^{\dagger}L_{\omega,\alpha}\rho\ln\rho] \nonumber \\
&=&-\sum_{\omega,\alpha,n,m} W^{\omega,\alpha}_{mn}p_{n}  \ln p_{m} + \sum_{\omega,\alpha,n,m} W^{\omega,\alpha}_{mn}p_{n}  \ln p_{n}. \label{dotS}
\eeqa
By combining Eqs.~(\ref{dotQdet}) and (\ref{dotS}), the second law of thermodynamics can be shown from the nonnegativity of the relative entropy
\beq
\dot{\sigma}=\sum_{\omega,\alpha,n,m}\hspace{-3mm}{}^{'}W^{\omega,\alpha}_{mn}p_{n}\ln\frac{W^{\omega,\alpha}_{mn}p_{n}}{W^{-\omega,\alpha}_{nm}p_{m}}\geq 0.\label{SL}
\eeq
Here, we note that the expression~(\ref{SL}) is similar to that of the classical entropy production rate, and $W^{\omega,\alpha}_{mn}$ can be interpreted as a quantum counterpart of the transition rate matrix. The nonnegativity of the entropy production implies that the Lindblad master equation provides thermodynamically consistent dynamics, and hence one may discuss the mechanism of QSL in thermodynamic viewpoint based on this dynamics.
In this direction, we utilize the technique used in Ref.~\cite{Shiraishi18} and relate the speed of the quantum evolution with the entropy production as we do in the next section.

%Then, the entropy production rate $\dot{\sigma}:=\dot{S}-\beta \dot{Q}$ is given by
%\beqa
%\dot{\sigma}&=&\sum_{\omega,\alpha,n,m}W^{\omega,\alpha}_{mn}p_{n}\ln\frac{W^{\omega,\alpha}_{mn}p_{n}}{W^{-\omega,\alpha}_{nm}p_{m}}\nonumber \\
%&=&\frac{1}{2}\sum_{\omega,\alpha,n,m}\Bigl(W^{\omega,\alpha}_{mn}p_{n}-W^{-\omega,\alpha}_{nm}p_{m}\Bigr)\ln\frac{W^{\omega,\alpha}_{mn}p_{n}}{W^{-\omega,\alpha}_{nm}p_{m}}.
%\eeqa

%Note that in the classical limit, the density matrix commutes with the Hamiltonian of the system. Thus, $W^{kl}_{mn}=W^{\text{cl}}_{mn}\delta_{kn}\delta_{lm}$. The classical limit of the Lindblad equation reproduces the Master equation as follows:
%\beq
%\partial_{t}p_{m}=\langle m|\partial_{t}\rho|m\rangle = \sum_{k,l,n}\left(W^{kl}_{mn}p_{n}- W^{kl}_{nm}p_{m}\right)\longrightarrow \sum_{n}\left(W^{\text{cl}}_{mn}p_{n}-W^{\text{cl}}_{nm}p_{m}\right). \label{EQOMform}
%\eeq
 %Moreover, we recover the classical heat: $
%-\beta\dot{Q}\longrightarrow \beta\sum_{n,m}p_{n}W^{\text{cl}}_{mn}\Delta_{n,m}=\sum_{mn}p_{n}W^{\text{cl}}_{mn}(E_{n}-E_{m}).$

\section{\label{sec:derivation}Derivation of the quantum speed limit}

\subsection{Quantum speed limit and the decomposition of the generator of the time-evolution}

In this paper, we now choose the trace norm
\beq
||X||_{\rm{tr}}:=\frac{1}{2}\Tr\Bigl[\sqrt{X^{\dagger}X}\Bigr]
\eeq
and also introduce the trace distance 
\beq
T(\rho,\rho'):=||\rho-\rho'||_{\rm{tr}},
\eeq
because this choice allows us to relate the norm of the generator to the quantities that appear in stochastic thermodynamics. Therefore, by using~(\ref{GOMSL}), we start from the formal inequality
\beq
\tau\geq \frac{T(\rho(0),\rho(\tau))}{\av{||\dot{\rho}||_{\rm{tr}}}_{\tau}}. \label{TrQSL}
\eeq
Our aim in this paper is to further bound the right hand side in Eq.~(\ref{TrQSL}) from below by several physical quantities, so that one can extract physical mechanism determining the speed in quantum dynamics. 

%From Eq.~(\ref{TrQSL}), we find that the speed of the state transformation is fully determined by the norm of the generator of the time-evolution, i.e., $||\dot{\rho}||_{\rm{tr}}=||\mathcal{L}_{t}[\rho]||_{\rm{tr}}$. 
In the following, we would like to focus on the velocity term $||\dot{\rho}||_{\rm{tr}}$ and discuss how the generator $\mathcal{L}_{t}$ will change in time the diagonal element $p_{n}(t)$ and the eigenbasis $|n(t)\rangle$ of the density matrix $\rho(t)=\sum_{n}p_{n}(t)|n(t)\rangle\langle n(t)|$. For this purpose, we split the dissipater as $\mathcal{D}[\rho]=\mathcal{D}_{\rm{d}}[\rho]+\mathcal{D}_{\rm{nd}}[\rho]$, where 
%For later convenience, we rewrite the master equation as follows:
%\beq
%\partial_{t}\rho=-\frac{i}{\hbar}[H+H_{\rm{LS}},\rho]+\mathcal{D}_{\rm{d}}[\rho]+\mathcal{D}_{\rm{nd}}[\rho],
%\eeq
%where we split the dissipater into the diagonal part 
\beq
\mathcal{D}_{\rm{d}}[\rho]:=\sum_{n}\langle n|\mathcal{D}[\rho(t)]|n\rangle |n\rangle\langle n|=\sum_{n}\partial_{t}p_{n}|n\rangle\langle n|\label{Dissidiag}
\eeq
is the diagonal part and
\beq
\mathcal{D}_{\rm{nd}}[\rho]:=\sum_{m\neq n}\langle m|\mathcal{D}[\rho(t)]|n\rangle |m\rangle\langle n| \label{Dissindiag}
\eeq
is the non-diagonal part of the dissipater with respect to the eigenbasis $|n(t)\rangle$, respectively. By using the triangle inequality, we have
\beq
||\dot{\rho}||_{\rm{tr}}=||\mathcal{L}_{t}[\rho]||_{\rm{tr}}\leq \frac{1}{\hbar}||[H,\rho]||_{\rm{tr}}+||\mathcal{D}_{\rm{d}}[\rho]||_{\rm{tr}}+||\mathcal{D}_{\rm{nd}}[\rho]||_{\rm{tr}}. \label{Trdotrho}
\eeq
Here, the first term on the right-hand side of Eq.~(\ref{Trdotrho}) characterizes the speed of the unitary time-evolution generated by the system Hamiltonian, as is the case for isolated quantum systems. From Eq.~(\ref{Dissidiag}), we find that the second term
\beq
||\mathcal{D}_{\rm{d}}[\rho]||_{\rm{tr}}=\sum_{n}|\partial_{t}p_{n}|
\eeq
characterizes the speed of the population transfer. Finally, we consider the third term $||\mathcal{D}_{\rm{nd}}[\rho]||_{\rm{tr}}$. By introducing the Hermitian operator~\cite{footnote}
\beq
H_{\mathcal{D}}(t):=\sum_{m\neq n}\frac{i\hbar \langle m|\mathcal{D}[\rho(t)]|n\rangle }{p_{n}-p_{m}} |m\rangle \langle n|, \label{DissiH}
\eeq
we can rewrite Eq.~(\ref{Dissindiag}) as
\beq
\mathcal{D}_{\rm{nd}}[\rho]=-\frac{i}{\hbar}[H_{\mathcal{D}}(t),\rho(t)]. \label{oddissiH}
\eeq
We then find that $\mathcal{D}_{\rm{nd}}$ describes part of the bath dynamics which generates a unitary time-evolution. %Therefore, both $H$ and $H_{\mathcal{D}}$ are related to the speed of changing the eigenstates $|n(t)\rangle$. 
In the following, we would like to discuss the geometric meaning of $H$ and $H_{\mathcal{D}}$. In doing so, we introduce the following quantity
\beq
\xi(t):=\sum_{m\neq n}\frac{i\hbar\langle m| \partial_{t}\rho |n\rangle}{p_{n}-p_{m}}|m\rangle\langle n|, \label{nonadH}
\eeq
which is a generator to escort the state along $|n(t)\rangle$ as
\beq
|n(t)\rangle \rightarrow \left(1+\frac{\delta t}{i\hbar}\xi(t)\right)|n(t)\rangle = e^{i\delta t A_{n}(t)}|n(t+\delta t)\rangle,
\eeq
and $A_{n}(t):=i\langle n(t)|\partial_{t}n(t)\rangle$ is the Berry connection. This means $\xi$ transports the state along the same label $n$ of the non-adiabatic state $|n(t)\rangle$, i.e., $\xi$ generates a parallel transport and $(1/i\hbar)\xi$ can be interpreted as a geometric connection. Now if we use the Lindblad master equation~(\ref{LE}), we can show that 
\beq
\xi=\tilde{H}+H_{\mathcal{D}}, \label{nonadHD}
\eeq
where $\tilde{H}:=H-\sum_{n}|n\rangle\langle n|H|n\rangle\langle n|$. We therefore respectively interpret $H$ and $H_{\mathcal{D}}$ as the Hamiltonian part and the bath part of the non-adiabatic geometric connection $\xi$ for Lindblad dynamics [note that the actions of $H$ and $\tilde{H}$ on $|n(t)\rangle$ give only an irrelevant U(1) phase difference]. 
%Although $H_{\mathcal{D}}$ is somewhat abstract at this stage, 
It is shown later that $H_{\mathcal{D}}$ reproduces the counter-diabatic Hamiltonian~\cite{STAR,DR03,DR05,Berry09} which generates a parallel transport for the adiabatic energy eigenstate $|\epsilon_{n}(t)\rangle$, when the speed of the external driving is very slow and the thermal relaxation is dominant.

%\beq
%\Tr|\partial_{t}\rho|\leq  \sum_{m}|\partial_{t}p_{m}|+\Tr\Bigl| -\frac{i}{\hbar}[H(t),\rho(t)]\Bigr|+\Tr\Bigl| -\frac{i}{\hbar}[H^{\mathcal{D}}(t),\rho(t)]\Bigr|, \label{Traceineq}
%\eeq
%and bound from above the three terms on the right-hand side of Eq.~(\ref{Traceineq}) as follows.

\subsection{Bounding the norm of generators}
Having identified how each term in Eq.~(\ref{Trdotrho}) changes $p_{n}(t)$ and $|n(t)\rangle$ in time, we now relate those terms with the energy fluctuation and the entropy production as follows.

We first bound from above the term $\hbar^{-1}||[H,\rho]||_{\rm{tr}}$ as follows.  
We note that the trace norm is contractive under a completely-positive and trace-preserving (CPTP) map $\Phi$~\cite{Nielsen}:
\beq
||\Phi(X)||_{\rm{tr}}\leq ||X||_{\rm{tr}}. \label{contractive}
\eeq
Now let us denote $|\rho\rangle$ as the purification of $\rho$ and take $\Phi$ as the partial trace $\Phi(|\rho\rangle\langle \rho|)=\rho$.  We also introduce a natural extension of $H$ to a bigger space, denoted by $\bar{H}$, satisfying $\Phi(\bar{H})=H$. We apply~(\ref{contractive}) and obtain
\beqa
\frac{1}{\hbar}||[H,\rho]||_{\rm{tr}} &\leq & \frac{1}{\hbar} || [\bar{H},|\rho\rangle\langle \rho|] ||_{\rm{tr}}\nonumber \\
%&=&\frac{1}{2\hbar}\Tr\sqrt{\left(\bar{H}-\langle\rho|\bar{H}|\rho\rangle\right)|\rho\rangle\langle\rho|\left(\bar{H}-\langle\rho|\bar{H}|\rho\rangle\right)+(\Delta E)^{2}|\rho\rangle\langle\rho|} \nonumber \\
&=&\frac{\Delta E}{2\hbar}\Tr \left[\sqrt{|\rho_{\perp}\rangle\langle\rho_{\perp}|+|\rho\rangle\langle\rho|}\right] \nonumber \\
&=&\frac{1}{\hbar}\Delta E, \label{Hresult}
\eeqa
where
\beq
(\Delta E)^{2}:=\langle\rho|\bar{H}^{2}|\rho\rangle-\langle\rho|\bar{H}|\rho\rangle^{2}=\Tr[H^{2}\rho]-(\Tr[H\rho])^{2}
\eeq
is the energy fluctuation and
\beq
|\rho_{\perp}\rangle:=\frac{\left(\bar{H}-\langle\rho|\bar{H}|\rho\rangle\right)|\rho\rangle}{\Delta E}
\eeq
is a state which is orthogonal to $|\rho\rangle$. As a result, we can bound from above the norm of the generator induced by the Hamiltonian by the energy fluctuation~(\ref{Hresult}).

We next bound from above the term $||\mathcal{D}_{\rm{nd}}[\rho]||_{\rm{tr}}$. From Eq.~(\ref{oddissiH}), we can use a method similar to that given in Eq.~(\ref{Hresult}) and obtain  
\beq
||\mathcal{D}_{\rm{nd}}[\rho]||_{\rm{tr}}=\frac{1}{\hbar}||[H_{\mathcal{D}},\rho]||_{\rm{tr}}\leq \frac{1}{\hbar}\Delta E_{\mathcal{D}}, \label{odresult}
\eeq
where 
\beq
(\Delta E_{\mathcal{D}})^{2}:=\Tr[H_{\mathcal{D}}^{2}\rho]
\eeq
is the fluctuation of the bath part $H_{\mathcal{D}}$ of the generator $\xi$~(\ref{nonadHD}). %which generates a parallel transport of the non-adiabatic state $|n(t)\rangle$. 
Since the fluctuation of the Hamiltonian part $H$ of $\xi$, i.e., the energy fluctuation $\Delta E$, is interpreted as the velocity for isolated systems, we interpret $\Delta E_{\mathcal{D}}$ as the velocity of the bath induced unitary dynamics. 
%is the fluctuation of the Hermitian operator $H_{\mathcal{D}}$ which generates the unitary part of the time-evolution~(\ref{oddissiH}) induced by the bath. 
Here, note that $\Tr[H_{\mathcal{D}}\rho]=0$.

We finally bound from above the term $||\mathcal{D}_{\rm{d}}[\rho]||_{\rm{tr}}$ and relate it to the quantities which appear in stochastic thermodynamics. Note that the time derivative of $p_{m}$ satisfies the classical master equation-like relation
\beq
\partial_{t}p_{m}=\langle m|\mathcal{D}[\rho]|m\rangle=\sum_{\omega,\alpha,n}\hspace{-1.2mm}{}^{'}\left(W^{\omega,\alpha}_{mn}p_{n}-W^{-\omega,\alpha}_{nm}p_{m}\right). \label{diagmaster}
\eeq
We then have
\beqa
||\mathcal{D}_{\rm{d}}[\rho]||_{\rm{tr}}&=&\frac{1}{2}\sum_{m}\left|\sum_{\omega,\alpha,n}\hspace{-1.2mm}{}^{'}\left(W^{\omega,\alpha}_{mn}p_{n}-W^{-\omega,\alpha}_{nm}p_{m}\right) \right| \nonumber \\
&\leq & \frac{1}{2}\sum_{m}\sqrt{\sum_{\omega,\alpha,n}\hspace{-1.2mm}{}^{'}\frac{(W^{\omega,\alpha}_{mn}p_{n}-W^{-\omega,\alpha}_{nm}p_{m})^{2}}{W^{\omega,\alpha}_{mn}p_{n}+W^{-\omega,\alpha}_{nm}p_{m}}\sum_{\omega,\alpha,n}\hspace{-1.2mm}{}^{'}\left(W^{\omega,\alpha}_{mn}p_{n}+W^{-\omega,\alpha}_{nm}p_{m}\right) } \nonumber \\
&\leq &\frac{1}{2}\sqrt{\sum_{\omega,\alpha,n, m}\hspace{-3mm}{}^{'}\frac{(W^{\omega,\alpha}_{mn}p_{n}-W^{-\omega,\alpha}_{nm}p_{m})^{2}}{W^{\omega,\alpha}_{mn}p_{n}+W^{-\omega,\alpha}_{nm}p_{m}}\sum_{\omega,\alpha,n, m}\hspace{-3mm}{}^{'}\left(W^{\omega,\alpha}_{mn}p_{n}+W^{-\omega,\alpha}_{nm}p_{m}\right) } \nonumber \\
&\leq & \sqrt{\frac{\dot{\sigma}A}{2}},  \label{Entineq}
\eeqa
where we use the Cauchy-Schwartz inequality twice and obtain the second and the third line. In deriving the last line of~(\ref{Entineq}), we use the following inequality
\beq
\sum_{\omega,\alpha,n, m}\hspace{-3mm}{}^{'}\frac{(W^{\omega,\alpha}_{mn}p_{n}-W^{-\omega,\alpha}_{nm}p_{m})^{2}}{W^{\omega,\alpha}_{mn}p_{n}+W^{-\omega,\alpha}_{nm}p_{m}}\leq\frac{1}{2} \sum_{\omega,\alpha,n,m}\hspace{-3mm}{}^{'}(W^{\omega,\alpha}_{mn}p_{n}-W^{-\omega,\alpha}_{nm}p_{m})\ln\frac{W^{\omega,\alpha}_{mn}p_{n}}{W^{-\omega,\alpha}_{nm}p_{m}}=\dot{\sigma},
\eeq
which follows from $2(a-b)^{2}/(a+b)\leq (a-b)\ln(a/b) $ for nonnegative $a$ and $b$. We also introduce the quantum dynamical activity which is analogous to the classical dynamical activity~\cite{LAW,Gar,Maes1,Maes2,SST} as
\beq
A:=\frac{1}{2}\sum_{\omega,\alpha,n, m}\hspace{-3mm}{}^{'}\left(W^{\omega,\alpha}_{mn}p_{n}+W^{-\omega,\alpha}_{nm}p_{m}\right). \label{activity}
\eeq
%where the summation in~(\ref{activity}) should exclude indicies satisfying $W^{\omega,\alpha}_{mn}p_{n}=W^{-\omega,\alpha}_{nm}p_{m}$, i.e., $(m=n)\wedge (\omega=0)$. 
Here, Eq.~(\ref{activity}) quantifies how frequently the jumps between different $p_{m}$'s occur, as can be checked by comparing it with the classical master equation-like relation~(\ref{diagmaster}). We note that inequality~(\ref{Entineq}) is essentially the same as the one used in Ref.~\cite{Shiraishi18} to derive the classical speed limit for stochastic processes. However, we emphasize that $\dot{\sigma}$ and $A$ appearing in~(\ref{Entineq}) are fully quantum and they differ from their classical counterparts.

\subsection{Main result}

Combining Eqs.~(\ref{Hresult}), (\ref{odresult}) and (\ref{Entineq}), we finally obtain
\beq
||\partial_{t}\rho||_{\rm{tr}} \leq \frac{1}{\hbar}\Delta E+\frac{1}{\hbar}\Delta E_{\mathcal{D}}+\sqrt{\frac{1}{2}\dot{\sigma}A}. \label{dtrhoineq}
\eeq
Here, the first and the second terms are related to the speed of changing $|n(t)\rangle$ by the unitary time-evolution generated by the Hamiltonian and the dissipater, respectively. The third term is related to the speed of changing $p_{n}(t)$ induced by the bath.  %By taking the time-integral of Eq.~(\ref{dtrhoineq}) and using~(\ref{TrQSL}), we obtain our main result, the QSL for open systems:
We now take the time-integral of Eq.~(\ref{dtrhoineq}) and further bound from above the right-hand side by using the Cauchy-Schwartz inequality $\langle\sqrt{\dot{\sigma}A}\rangle_{\tau}\leq \sqrt{\langle\dot{\sigma}\rangle_{\tau}\langle A\rangle_{\tau}}$. By combining it with~(\ref{TrQSL}), we obtain our main result, the QSL for open systems:
%\beq
%\tau\geq \frac{T(\rho(0),\rho(\tau))}{\frac{1}{\hbar}\langle \Delta E\rangle_{\tau}+\langle ||\mathcal{D}_{\rm{nd}}[\rho]||_{\rm{tr}}\rangle_{\tau}+\sqrt{\frac{1}{2}\langle\dot{\sigma}\rangle_{\tau}\langle A\rangle_{\tau}}}. \label{RESULTa}
%\eeq
%In the following, we discuss the term $||\mathcal{D}_{\rm{nd}}[\rho]||_{\rm{tr}}$. To best of our knowledge, we could not find a direct connection of this term with a well-known physically meaningful quantity. 
\beq
\tau\geq\frac{T(\rho(0),\rho(\tau))}{\hbar^{-1}\langle \Delta E \rangle_{\tau}+\hbar^{-1}\langle \Delta E_{\mathcal{D}}\rangle_{\tau}+\sqrt{\frac{1}{2}\langle\dot{\sigma}\rangle_{\tau}\langle A\rangle_{\tau}}} . \label{RESULT}
\eeq
Here, the operation time $\tau$ is bounded from below by the trace distance $T(\rho(0),\rho(\tau))$ divided by the sum of three different average velocity terms $\hbar^{-1}\langle \Delta E \rangle_{\tau}$, $\hbar^{-1}\langle \Delta E_{\mathcal{D}}\rangle_{\tau}$ and $\sqrt{\langle\dot{\sigma}\rangle_{\tau}\langle A\rangle_{\tau}/2}$. The average energy fluctuation $\hbar^{-1}\langle \Delta E \rangle_{\tau}$ characterizes the speed of the state transformation via $H$ as is the case for isolated quantum systems. If we want to speed up the state transformation, we have to increase the intensity of the Hamiltonian and thus large energy fluctuation is required. The combination of the entropy production and the dynamical activity $\sqrt{\langle\dot{\sigma}\rangle_{\tau}\langle A\rangle_{\tau}/2}$ characterizes the speed of the population transfer via the bath, as is the case for classical stochastic systems. By increasing the strength of the effect of the bath, we can speed up the population transfer but large entropy production and dynamical activity are required. The term $\hbar^{-1}\langle \Delta E_{\mathcal{D}}\rangle_{\tau}$ is related to the non-adiabatic geometric connection of the Lindblad dynamics, and characterizes the speed of the unitary evolution induced by the bath. 
%is related to the speed of the unitary rotation part of the dynamics induced by the bath. 
We note that this term has no counterpart in previously obtained classical speed limit inequality~\cite{Shiraishi18}, since the bath only induces population transfer in the classical regime. We therefore emphasize that this term is needed to estimate or to obtain intuition about the minimum operation time $\tau$ in open quantum systems. In Sec.~\ref{sec:strongdissipation}, we further discuss the physical meaning of this term by considering the thermodynamically quasi-adiabatic regime in which the thermal relaxation becomes dominant.  We also discuss how the obtained QSL~(\ref{RESULT}) reproduces the speed limit for isolated quantum systems and for classical stochastic systems in the next section.

\section{\label{sec:limit}Limiting cases of the quantum speed limit}
In this section, we consider three limiting cases of the quantum speed limit~(\ref{RESULT}); the dissipationless (quantum isolated system) limit, the thermodynamically quasi-adiabatic regime and the classical limit. Let us introduce the typical time-scale of the system Hamiltonian $H(t)$ as $\tau_{\rm{S}}$ and that for the  dissipater $\mathcal{D}$ induced by the bath as $\tau_{\rm{B}}$. We then introduce a parameter $\lambda=\tau_{\rm{S}}/\tau_{\rm{B}}$ and rewrite the Lindblad master equation as
\beq
\partial_{t}\rho(t)=-\frac{i}{\hbar}[H(t),\rho(t)]+\lambda \mathcal{D}[\rho(t)] , \label{expandLE}
\eeq
by rescaling the quantities $t$, $H$, and $\mathcal{D}$ appropriately.

\subsection{Dissipationless limit}
In the case of $\tau_{\rm{S}}\ll\tau_{\rm{B}}\ (\rightarrow \infty)$, i.e., $\lambda\rightarrow 0$, we can treat the system as isolated. In this case, only the energy fluctuation of the system is relevant and Eq.~(\ref{RESULT}) reproduces the Mandelstam-Tamm-type quantum speed limit
\beq
\tau\geq\frac{ T(\rho(0),\rho(\tau))}{\hbar^{-1}\langle \Delta E \rangle_{\tau}}.\label{QSLisolatedlimit}
\eeq
Note that the distance used in Eq.~(\ref{QSLisolatedlimit}) is not the Bures distance but the trace distance, and hence this formula should be regarded as a variant of the standard Mandelstam-Tamm relation.

\subsection{\label{sec:strongdissipation} Thermodynamically quasi-adiabatic regime}
 
Next, let us consider the thermodynamically quasi-adiabatic regime in which the thermal relaxation becomes dominant compared to the unitary dynamics of the system: $\tau_{\rm{B}}\ll\tau_{\rm{S}}$, i.e., $\lambda\gg 1$. In this regime, the system is quickly thermalized by the bath, and the density matrix of the system becomes close to the instantaneous stationary state $\rho_{\rm{eq}}(t)$. 

Therefore, we expand the density matrix in the series of $\lambda^{-1}$ as
\beq
\rho(t)=\rho_{\rm{eq}}(t)+\lambda^{-1}\delta\rho^{(1)}(t)+\lambda^{-2}\delta\rho^{(2)}(t)+\cdots. \label{expandrho}
\eeq
We substitute Eq.~(\ref{expandrho}) into Eq.~(\ref{expandLE}) and obtain
\beq
\partial_{t}\rho_{\rm{eq}}(t)+\lambda^{-1}\partial_{t}\delta\rho^{(1)}(t)+\cdots =-\frac{i}{\hbar}[H(t),\lambda^{-1}\delta\rho^{(1)}(t)+\cdots]+\mathcal{D}[\delta\rho^{(1)}(t)+\lambda^{-1}\rho^{(2)}(t)+\cdots]. \label{expandLEa}
\eeq
By comparing the lowest order $\lambda^{-1}$ terms on both hand sides of Eq.~(\ref{expandLEa}), we have
\beq
\partial_{t}\rho_{\rm{eq}}(t)=\mathcal{D}[\delta\rho^{(1)}(t)]+O(\lambda^{-1}). \label{LOD}
\eeq

Let us write the instantaneous Gibbs distribution in the representation diagonalizing the Hamiltonian as $\rho_{\rm{eq}}(t)=\sum_{n}p^{\rm{eq}}_{n}(t)|\epsilon_{n}(t)\rangle\langle \epsilon_{n}(t)|$. %and the energy eigenstate at time $t$ as $p^{\rm{eq}}_{n}(t)$ and $|\epsilon_{n}(t)\rangle$, respectively, i.e., $\rho_{\rm{eq}}(t)=\sum_{n}p^{\rm{eq}}_{n}(t)|\epsilon_{n}(t)\rangle\langle \epsilon_{n}(t)|$. 
Then, by using Eq.~(\ref{LOD}) and noting that $p_{n}(t)=p^{\rm{eq}}_{n}(t)+O(\lambda^{-1})$ and $|n(t)\rangle=|\epsilon_{n}(t)\rangle+O(\lambda^{-1})$, $H_{\mathcal{D}}(t)$ given in Eq.~(\ref{DissiH}) can be approximated as
\beq
H_{\mathcal{D}}(t)=H_{\rm{cd}}(t)+O(\lambda^{-1}),
\eeq
where 
\beqa
H_{\rm{cd}}(t)&:=&i\hbar\sum_{n}(1-|\epsilon_{n}(t)\rangle\langle \epsilon_{n}(t)|)|\partial_{t}\epsilon_{n}(t)\rangle\langle \epsilon_{n}(t)| \nonumber \\
 &=& \sum_{m\neq n} \frac{i\hbar\langle \epsilon_{m}|\partial_{t}H|\epsilon_{n}\rangle }{\epsilon_{n}-\epsilon_{m}}|\epsilon_{m}\rangle \langle \epsilon_{n}|    
\eeqa
is the counter-diabatic Hamiltonian which enforces the state to follow the instantaneous energy eigenstate $|\epsilon_{n}(t)\rangle$ for an isolated quantum system to realize the shortcuts to adiabaticity protocol~\cite{STAR,DR03,DR05,Berry09}. Also, $\Delta E_{\mathcal{D}}$ satisfies 
\beq
(\Delta E_{\mathcal{D}})^{2}=(\Delta E_{\rm{cd}})^{2}+O(\lambda^{-2})=\hbar^{2}\sum_{n}p^{\rm{eq}}_{n}(t)g^{n}_{\rm{FS}}+O(\lambda^{-2}),
\eeq 
%Then, from Eq.~(\ref{LOD}), we have
%\beqa
%||\mathcal{D}_{\rm{nd}}[\rho]||_{\rm{tr}}&=&\Bigl|\Bigl|\sum_{n}p_{n}^{\rm{eq}}(|\partial_{t}\epsilon_{n}\rangle \langle\epsilon_{n}|+|\epsilon_{n}\rangle \langle\partial_{t}\epsilon_{n}|)\Bigr|\Bigr|_{\rm{tr}} \nonumber \\ 
%&\leq&\sum_{n}p_{n}^{\rm{eq}} ||(|\partial_{t}\epsilon_{n}\rangle \langle\epsilon_{n}|+|\epsilon_{n}\rangle \langle\partial_{t}\epsilon_{n}|)||_{\rm{tr}} \nonumber \\
%&=&\sum_{n}p_{n}^{\rm{eq}}\sqrt{g_{\rm{FS}}^{n}} \nonumber \\
%&\leq &\sqrt{\sum_{n}p_{n}^{\rm{eq}}g_{\rm{FS}}^{n}}. \label{ndboundSD}
%\eeqa
where $(\Delta E_{\rm{cd}})^{2}:=\Tr[H_{\rm{cd}}^{2}(t)\rho_{\rm{eq}}(t)]$ is the energy fluctuation measured in terms of the counter-diabatic Hamiltonian and
\beq
g_{\rm{FS}}^{n}:=\langle \partial_{t}\epsilon_{n}(t)|(1-|\epsilon_{n}(t)\rangle\langle \epsilon_{n}(t)|)|\partial_{t}\epsilon_{n}(t)\rangle \label{FSmetric}
\eeq
is the Fubini-Study metric~\cite{Gibbons} which gives the curvature of the $|\epsilon_{n}(t)\rangle$-state manifold. We note that Eq.~(\ref{FSmetric}) measures the quadratic decay of the fidelity between two neighboring states: $|\langle \epsilon_{n}(t)|\epsilon_{n}(t+dt)\rangle|^{2}=1-g^{n}_{\rm{FS}}dt^{2}+O(dt^{3})$.

From the above argument, the QSL~(\ref{RESULT}) becomes
\beq
\tau \geq %\frac{T(\rho(0),\rho(\tau))}{\langle \sqrt{\sum_{n}p_{n}^{\rm{eq}}g_{\rm{FS}}^{n}} \rangle_{\tau}+\sqrt{\frac{1}{2}\langle\dot{\sigma}\rangle_{\tau}\langle A\rangle_{\tau}}}=
\frac{T(\rho(0),\rho(\tau))}{\hbar^{-1}\langle \Delta E_{\rm{cd}} \rangle_{\tau}+\sqrt{\frac{1}{2}\langle\dot{\sigma}\rangle_{\tau}\langle A\rangle_{\tau}}}, \label{cdRESULT}
\eeq
when the speed of the external driving is very slow and the thermal relaxation is dominant. In this thermodynamically quasi-adiabatic regime, $\mathcal{D}_{\rm{nd}}$ enforces the basis $|n(t)\rangle$ which diagonalizes $\rho(t)$ to follow the instantaneous energy eigenbasis $|\epsilon_{n}(t)\rangle$. This energy eigenstate-tracking mechanism is similar to that of the shortcuts to adiabaticity protocol via the counter-diabatic Hamiltonian for an isolated system. Therefore, the counter-diabatic Hamiltonian $H_{\rm{cd}}$ or the geometry of the energy eigenstates $g^{n}_{\rm{FS}}$ becomes relevant for characterizing the speed of the state transformation induced by $\mathcal{D}_{\rm{nd}}$~\cite{footnote3}. 

\subsection{Classical limit}

We finally consider a classical limit of Eq.~(\ref{RESULT}) such that the dynamics is given by classical probabilistic processes. (Note that the classical limit here does not mean taking the limit $\hbar\to 0$.) We start by decomposing the Hamiltonian into the driven part $H_{1}(t)$ and the undriven part $H_{0}$ as $H(t)=H_{0}+H_{1}(t)$. In the classical limit, we require that (i) $[H_{0},H_{1}(t)]=0$, i.e., $H(t)=\sum_{n}\epsilon_{n}(t)|\epsilon_{n}\rangle\langle\epsilon_{n}|$ is also diagonalized with respect to the energy eigenstates $|\epsilon_{n}\rangle$ of the undriven Hamiltonian $H_{0}$, and (ii) the initial density matrix of the system has no coherence in the energy eigenbasis $|\epsilon_{n}\rangle$, i.e., $\rho(0)=\sum_{n}p_{n}(0)|\epsilon_{n}\rangle\langle\epsilon_{n}|$. Let us also assume that the spectrum of the undriven Hamiltonian $H_{0}$ is non-degenerate. Then, from Eq.~(\ref{LE}), we find that the population $P_{n}(t):=\langle\epsilon_{n}|\rho(t)|\epsilon_{n}\rangle$ of the eigenbasis $|\epsilon_{n}\rangle$ satisfies the Pauli master equation~\cite{Breuer}  
\beq
\partial_{t}P_{n}(t)=\sum_{\alpha,m}\left(M^{\alpha}_{nm}(t)P_{m}(t)-M^{\alpha}_{mn}(t)P_{n}(t)\right), \label{CLME}
\eeq
where 
\beq
M^{\alpha}_{nm}(t):=\gamma_{\alpha}(\epsilon_{m}(t)-\epsilon_{n}(t))|\langle\epsilon_{n}|L_{k}|\epsilon_{m}\rangle|^{2}
\eeq
reproduces the classical transition rate matrix from $n$ to $m$ in the classical limit. The off-diagonal component $\rho_{mn}(t):=\langle \epsilon_{m}|\rho(t)|\epsilon_{n}\rangle$ of the density matrix satisfies
\beq
\partial_{t}\rho_{mn}(t)=-\frac{i}{\hbar}(\epsilon_{m}(t)-\epsilon_{n}(t))\rho_{mn}(t)-\frac{1}{2}\sum_{\alpha,l}\left(M^{\alpha}_{lm}(t)+M^{\alpha}_{ln}(t)\right)\rho_{mn}(t),
\eeq
and from the requirement (ii), $\rho_{mn}(t)=0$ for all $t$.

We therefore find that in the classical limit (i) and (ii), we have $p_{n}(t)\rightarrow P_{n}(t)$, $|n(t)\rangle\rightarrow |\epsilon_{n}\rangle$, $W^{\omega_{t},\alpha}_{mn}\rightarrow M^{\alpha}_{mn}(t)\delta(\omega_{t}-\epsilon_{n}(t)+\epsilon_{m}(t))$, and Eq.~(\ref{CLME}) reproduces the classical master equation. 
%Let us use the energy eigenbasis to describe the density matrix and consider the phase factor of the off-diagonal components induced by the Hamiltonian. In the classical limit, the phase  rapidly oscillates and thus the off-diagonal components vanish. Therefore, $\rho(t)$ is diagonalized with respect to the energy eigenbasis $|\epsilon_{n}(t)\rangle$ in the classical limit and thus Eq.~(\ref{QMtrans}) becomes $W^{\omega,\alpha}_{m,n}=M^{k}_{m,n}\delta(\omega-\epsilon_{n}+\epsilon_{m})$ where $M^{k}_{m,n}$ is the classical transition rate matrix describing the transition $n\rightarrow m$. As a result, Eq.~(\ref{diagmaster}) reproduces the classical master equation
In addition, the entropy production~(\ref{EPrate}) and the dynamical activity~(\ref{activity}) reproduce their classical counterparts. Since $\rho(t)$ is diagonalized with respect to the energy eigenbasis $|\epsilon_{n}\rangle$, we have $||[H,\rho]||_{\rm{tr}}=0$ and $||\mathcal{D}_{\rm{nd}}[\rho]||_{\rm{tr}}=0$. We further note that the trace distance reproduces the total variational distance as 
\beq
 T(\rho(0),\rho(\tau))=\frac{1}{2}\sum_{n}|p_{n}(0)-p_{n}(\tau)|. 
\eeq
Therefore, in the classical limit, we reproduce the classical speed limit for stochastic processes~\cite{footnote2}: 
\beq
 \tau\geq \frac{T(\rho(0),\rho(\tau))}{\sqrt{\frac{1}{2}\langle\dot{\sigma}\rangle_{\tau}\langle A\rangle_{\tau}}}.  \label{QSLclassical}
\eeq

\section{\label{sec:summary}Conclusion}

We have derived the QSL inequality~(\ref{RESULT}) for open quantum systems which generalizes the previously obtained inequalities for isolated quantum system and classical stochastic processes. By decomposing the generator of the time-evolution into three parts, we showed how each term changes the diagonal components and the eigenbasis of the density matrix in time. We then relate the norm of the decomposed generators to physically well-known quantities such as the energy fluctuation and the entropy production. This allows us to obtain better intuition about the speed in open quantum systems and the derived inequality should be relevant to various applications in quantum devices that are subject to decoherence and dissipation. Quite interestingly, when the external driving is slow and the thermal relaxation becomes dominant, the new velocity term which appears in open quantum systems is related to the counter-diabatic Hamiltonian used in shortcuts to adiabaticity. This relation may suggest further connection between finite-time quantum control theory and QSLs.

\begin{acknowledgments}
KF was supported by JSPS KAKENHI Grant Number JP18J00454. NS was supported by Grant-in-Aid for JSPS Fellows JP17J00393. KS was supported by JSPS Grants-in-Aid for Scientific Research (JP16H02211 and JP17K05587).
\end{acknowledgments}


\begin{thebibliography}{99}
\bibitem{footnote} If $p_{m}=p_{n}$ for some $m\neq n$, we relabel the indices as $m=(a,\alpha)$, $n=(b,\beta)$, etc. Then, Eq.~(\ref{DissiH}) should be replaced by $H_{\mathcal{D}}(t)=i\hbar \sum_{a\neq b,\alpha,\beta} ( p_{b}-p_{a} )^{-1}\langle a,\alpha|\mathcal{D}[\rho(t)]|b,\beta\rangle|a,\alpha\rangle \langle b,\beta|$.
\bibitem{footnote3}  For a general non-adiabatic dynamics, we can also derive the inequality
\beq
\tau \geq \frac{T(\rho(0),\rho(\tau))}{ \hbar^{-1}\langle \Delta \xi\rangle_{\tau}+\sqrt{\frac{1}{2}\langle\dot{\sigma}\rangle_{\tau}\langle A\rangle_{\tau}}},
\eeq
which is a non-adiabatic counterpart of Eq.~(\ref{cdRESULT}). Here, $\Delta \xi:=\Tr[\xi^{2}\rho]=\sum_{n}p_{n}\eta_{\rm FS}^{n}$ and $\eta_{\rm FS}^{n}:=\langle\partial_{t}n|(1-|n\rangle\langle n|)|\partial_{t}n\rangle$ is the Fubini-Study metric for $|n(t)\rangle$-state manifold. We also note that $\Delta \xi \leq \Delta E + \Delta E_{\mathcal{D}}$.
\bibitem{footnote2} We note that by taking the square of both hand sides of~(\ref{QSLclassical}), we have $\tau\geq L^{2}(p(0),p(\tau))/(2\Sigma\langle A\rangle_{\tau})$ with $\Sigma:=\int dt\dot{\sigma}$ and $L(p(0),p(\tau)):=\sum_{n}|p_{n}(0)-p_{n}(\tau)|$, which was reported in Ref.~\cite{Shiraishi18}.
\end{thebibliography}
\end{document}